# Macroquantum Effects in Astronomy

Prof. Alexander M. Ilyanok

*A great number of macroquantum laws connecting gravity and electromagnetism if found empirically. To describe them the model of anisotropic gravitational field is proposed. This field is build as a superposition of planes and force lines are changed by thin elastic discs. Interaction of gravitational objects is described by second order linear differential equation having in some limits Newton or Maxwell equations. Tending to similar description of electromagnetic and gravitational forces the Newton gravity law is complemented by dynamical part dependent on speed of relative movement of objects. Static part of the Newton law is provided by longitudinal waves. They penetrate along planes with the speed equal to $a^{-4}c = 3,526 \cdot 10^8 c$ where $c$ is the speed of light and $a$ is the fine structure constant.*

Quantum relativistic picture formed in the beginning of the 20[th] century had just met with a number of paradoxes. Thus, L. de Broglie [1] considered particles as wave packs moving with the group speed $v_g$ and wavelength

$$l = \frac{h}{m v_g} \qquad (1)$$

where $h$ is the Planck constant. He had supposed that there is some speed $v_f$ of wave connected to the group speed by the relation

$$v_g v_f = c^2. \qquad (2)$$

Equation (1) was considerably good fitted to some experiments. Because of that it had become as basic in quantum mechanics. From the other hand, equation (2) lead to the serious contradiction with special relativity because there always followed $v_f > c$ from it. To avoid contradictions with special relativity $v_f$ had become considered as phase speed which is not related to real interactions and can not be observed. Usually equation (2) was not taken into account as having no physical sense. But serious contradictions in equation (1) also were hidden. Thus, if $v_g \to 0$ so $\lambda \to \infty$ and wave pack length of a particle is expanding to the infinity. Or, otherwise, when $v_g \sim c$ the mass of a particle tends to the mass of the Earth, for example, and $\lambda \to 0$. So wavelength of the Earth is infinitely small. It is clear that the contradictions considered are out of sense. And it is necessary to eliminate them.

The second problem is connected with the theory of relativity. Developing theory of H. Lorentz and A. Poincare Einstein had stated that there was no ether in space and the light moved in vacuum with absolute speed $c$ independent on the speed of source [1]. At the next step Einstein had postulated that all physical laws are invariant under Lorentz transformations. Applying to the all-physical scene one may conclude that experimental facts obtained for relativistic properties of elementary particles may be expanded on condensed matter. But the serious barrier had appeared before Einstein – instantaneous interaction in the Newton's gravity law. In other words, there is no retarded factor in the Newton's law. Let us remind that to fit the Newton's law to the energy conservation law the retarded part is needed. Laplace was the first who had introduced this part and he had obtained that the speed of gravity interactions should be not less than $10^7 c$. It is in contradiction with special relativity.

Einstein, Abraham, Nordstrom and some of other researchers had tried to solve this contradiction by Lorentz invariant wave equations [2]. But they had failed to do that. So Einstein had refused to consider gravity as a field and had introduced a new type of ether − curved space-time. He had created general relativity theory. But in nonlinear space-time there are problems with conservation laws including gravity. Despite of contradictions of general relativity it has stimulated the development of new direction in theoretical physics based on the Riemann geometry − geometrophysics.

But all attempts to apply the models of geometrophysics to the real physical objects including to elementary particles had appeared completely non-constructive. Practically, general relativity could describe only two experimental facts that the Newton's theory did not describe. They are the anomaly shift of the Mercury perihelion and deviation of star rays by the Sun.

Thus, the internally contradictory quantum relativistic picture of world was created.

So let us return to the linear Newton's world. But let us mare the following suppositions. Gravity is described by the field. The speed of penetration of the interaction is finite but not restricted by the speed of light.

Let us consider a set of Lorentz transformations introducing new critical speeds connected with phase state of substance. For these groups Lorentz factor will as $\left[1-(v/c_n)^2\right]^{-1/2}$ where $c_n = \boldsymbol{b}\boldsymbol{a}^n c$, $n=\pm 0, 1, 2, 3, 4$ and $\boldsymbol{b}$ is a factor connected with type of movement of a particle with the speed $v$, $\boldsymbol{a}$ is the constant of fine structure, $c$ is the speed of light. Letter $c_n$ denotes the critical speeds of movement of a substance dependent on its phase state. When $n=0$ one obtains the ordinary Lorentz transformation. Let us note that in special and general relativity the formal transfer of laws found for elementary particles without spin is made for any phase state of substance. It is the sufficient methodical mistake because the spin of particles of substance is not taken into account. But the spin is sufficient in chemical bonds. As a consequence of that, a condensed matter can not move in vacuum with the speed exceeding $\boldsymbol{a} c (n=1)$ because it is simply evaporated.

Now let us consider a system of gravitationally connected objects like Sun system or the Galaxy as some type of macroscopic condensate. This condensate can not move in vacuum with the speed exceeding $\boldsymbol{a} c$. This fact is confirmed by a tremendous number of observations, for example, concerning the movement of galaxies relative to relict background radiation.

Let us assume that there is special speed of gravitational interaction between objects depending on limit speed of their relative motion ($n = 1, 2, 3, 4$) and the limit radius of interaction corresponds to each of them. When $n = -1, -2, -3, -4$ the speed of interaction is higher than speed of light. Let us show what is the matter of this unusual fact. If two bodies interact by the field between each other, so this field is a material object and, hence, it should possess some elasticity. In this framework one should differ the elastic model of the vacuum and the model of the field itself. In this case general wave equations in elastic medium may be used [3]:

$$\frac{\partial^2 \mathbf{u}}{\partial t^2} - c_l^2 \, graddiv \, \mathbf{u} + c_t^2 \, rotrot \, \mathbf{u} = \mathbf{F} \qquad (3)$$

where **u** is the vector of the field, **F** is the vector of special force external to the elastic medium, $c_t$ is the speed of transversal wave and $c_l$ is the speed of longitudinal wave. Newton's equations, Maxwell's equations and hydrodynamic equations are derived from this equation as partial cases.

In general, longitudinal and transversal waves in (3) are connected through the Poisson coefficient σ:

$$c_l = \frac{c_t \sqrt{2}}{\sqrt{1-2\sigma}} \qquad (4)$$

where for real elastic physical medium 0 ≤ σ < 0.5. If there is transversal wave in an elastic medium so it is always $c_l > \sqrt{2} c_t$.

Let us note that for the rubber or gel type media the coefficient σ≈0.5. In this case the speed of longitudinal wave exceeds the speed of transversal wave in many times. Let us consider a field as an elastic medium with σ≈0.5 and let us assume that the speed of transversal wave is less than the speed of light: $c_t ₤ c$.

In one's time Maxwell solving equation (3) could not determine the nature of longitudinal wave. He had obtained that the speed of electromagnetic wave was equal to the speed of light and, hence, the longitudinal speed exceeded the speed of light. Maxwell had commented this phenomenon very prudently. He wrote [4] that electromagnetism and optics are powerless when they are asked to confirm or to deny that longitudinal vibrations exist. To save longitudinal waves there should be an "electrical liquid" of definite density which parameters can not be obtained from the experiment. But subsequent researchers was constantly excluding the longitudinal wave from the consideration as a wave of no physical sense.

At the modern level of the development of physics the physical sense of Maxwell's longitudinal wave may be found. This is the type of waves that describes Coulomb and Newton forces. It is well known that elastic forces arisen in a round elastic disc depend on distance $R$ like $1/R^2$ and for bending of the plate they behave like $1/R^3$ in linear approximation [3]. Such forces describe the laws of interaction between charges and magnetic dipoles. That is, that the laws of Coulomb Ampere and Newton follow naturally from the model. They are the round layers that longitudinal waves penetrate down when σ≈0.5.

It was shown experimentally even in 19[th] century. At that time to interpret gravity Buerknes had shown that bodies floated on the surface of a liquid attracted when they pulsed in the same phase and rejected when they pulsed in opposite phases according the law $1/R^2$ [5]. Buerknes experiments demonstrate that only waves along the liquid surface realise exchange interaction.

So let us to consider that gravitational field of separate atoms on the space is not continuous but is discrete. It is consisted of 861 discrete layers − discs connected like an unfolded book. Our world felt as three-dimensional is formed by the superposition of two-dimensional spaces. In this world the penetration of waves along the layers is realised with the speed exceeded the speed of light when the movement of waves and elementary particles between layers can have the speed, in principle, lower than the speed of light.

In times of Newton and Coulomb elementary particles were unknown and nothing was known about connection between their masses and charges. For current image of world it necessary to introduce charges into the Newton gravity law and masses into the Coulomb law. In this case let us suppose that gravitational field as not totally compensated field of central symmetric dipole composed from positive and negative charges. Mossotti (1836) was the first who had proposed that[2].

To describe gravitational interactions like electromagnetic fields it is necessary to introduce equivalents of the scalar electrical and vector magnetic fields. At the first time it was noted by Heavisude. He had proved that gravitational field can be expressed, in principle, through electromagnetic one if one represents it as two connected fields [6].

This statement follows directly from the Helmholtz theorem: "Any vector field **U** if it is finite, one-valued and continuous can be represented by the sum **U**= grad ϕ + rot **A**» [3].

Let us state that for the force of interaction between objects following condition should yield: grad ϕ ≠0 and rot **A** ≠0. At the same time $c_t$ ≠0, $c_l$ ≠0 in (3). These conditions provide us an opportunity to conclude that any motion of the object of matter in its own field or in the field of the object interacting with it will be rectilinear-rotated. It is natural because we on the Earth move along a spiral round the Earth orbit, the Earth move spirally round the orbit of the Sun etc. Essentially, rectilinear motion can be attribute only to photons and neutrino. In this case $c$, in fact, is the critical speed of rectilinear motion. Let us note that elementary particles having the rest mass should spin around their axis. Taking into account all aforementioned let us represent the force of interaction between objects in general form:

$$F = ar^{-2} \pm br^{-3}, \qquad (5)$$

where the first term describes the static Coulomb part of force and the second term describes the dynamical magnetic component, connected with the movement of charges. Therefore, gravitational field should have the component connected with the mass movement.

So one should not consider a static gravitational field without taking into account the relative movement of masses. Usually, the movement of the satellite on the orbit is connected with the kinetic energy of its motion or centrifugal forces. But the centrifugal force can be interpreted as some "dynamical antigravity".

There is the serious problem in astronomy. It is in that the galaxies are rotating as linked co-rotating objects. But Newton and Einstein equations are not describing these movements. Because the second term in (5) is like 'magnetic' force it is proportional to the velocity of the object. This 'dynamical' part of the world gravity will appear, for instance, when one try to take into account the influence of the rotation of gravitational field with the Sun. The movement of the Solar 'wind' or movement of stars around the center of the Galaxy can be as demonstrations of the velocity dependent terms in gravitational force.

For example, for the Mercury the gravitational strength on its orbit with the radius $R_1$ will depend on equatorial speed of the Sun rotation $v_\Theta$ as following

$$g = \frac{GM_\Theta}{R_1^2}\left[1 + \frac{v_\Theta}{c}\frac{2R_\Theta}{R_1}\right], \qquad (6)$$

and the secular shift of the mercury perihelion will

$$\Delta w = 2p\sqrt{1+b} = 43{,}05'', \qquad (7)$$

where $b = \dfrac{v_\Theta}{c}\dfrac{2R_\Theta}{R_1} = 1.6002 \cdot 10^{-7}$ with mass and radius of the Sun $M_\Theta$ and $R_\Theta$. The experiment gives **Dw**=42,6″±0.9″.

Let us remind that stable movement of the planet perihelion the Newton had investigated. He had proved that the expression (5) is the sole case describing the stable motion with rotation of perihelion.

The other example of the action of the dynamical part in the modified Newton's law (6) is the deflection of light in the Solar gravitational field. Newton considered the light, as any other material body will take place in gravitational interactions. After 100 years Soldner (1802) had calculated this interaction. He had found that a star ray moving near the Sun should be deflected on the angle $q = 2GM_\Theta / R_\Theta c^2 \approx 0.83''$ [7]. After next 100 years it was found experimentally that the ray is really deflected by the Sun but the angle of deflection is in two times larger. It was theoretically described due to Einstein general relativity [2].

Let us show that our model gives the same results. Taking into account the dynamical part in (6) let us assume $v_\Theta = c$ — this is the speed of light moving near the Sun surface. As result the deflection of the ray is almost in two times more than in the result of Soldner. And it coincides with the results of general relativity and the experiment.

So to describe effects of anomaly motion of the Mercury perihelion and deflection of photons in the gravitational field of the Sun it is not necessary to appeal to the general relativity.

Let us return again to the problems of de Broglie. Let us introduce a generalized quantum of action in equation (1) $h_a = a^{-z}h$, in this z=1,2,3… is the scale power index. As result one obtains the generalized de Broglie equation for the discrete states of relative motion of interacting bodies expressed only through the electrodynamical constants:

$$l = \dfrac{h_a}{mc} = \dfrac{m_0 e^2}{2m a^{z+1}}, \qquad (8)$$

where *m* is the rest mass of the proton or electron, $m_0$ is the magnet constant from SI. Planck constant is, in general, not fundamental one but it is the combination of electromagnetic and world constants. With the help of equation (8) it is possible to connect the microworld and macroworld. It should not put the mass of condensed matter in the generalized de Broglie equation. The mass of elementary particles play the role here. So the de Broglie wavelength should be interpreted as the radius of the spiral of the object movement. It is related as to macro, as well as to microobjects because the structure of the proton and electron and their superposition are defined the type of their movement in macrospace.

Taking into account aforementioned notes let us suppose the Coulomb and gravitational interaction as the waves of a field connected with the mass and the charge. At that, the differential equation (3) has solutions describing stable motion only for layers connected with the step $a^{-1}$=137,0360547255. This is the constant of fine structure and it has found from purely geometrical properties by the formula, shown in the Table 1, i. 1. Solutions of the equation (3) with definite boundary conditions for different cosmic bodies are represented in Table 1. The comparison of theoretical and experimental results one may conclude that the generalized de Broglie equation directly connects micro and macroworld.

Let us consider the second de Broglie equation (2). Let us put the transversal wave into correspondence to the group velocity and let us suppose that the relative motion of the objects is connected with this kind of wave. So the longitudinal wave is connected with the phase velocity. In this case vector product of velocities of longitudinal and transversal waves in plate elastic medium (field) is defined by the inequality:

$$|\boldsymbol{a}^{-n}\mathbf{c}_l \times \boldsymbol{a}^{+n}\mathbf{c}_t| \leq \mathbf{c}^2 \text{ or } |\mathbf{c}_l \times \mathbf{c}_t| \leq \mathbf{c}^2 \qquad (9)$$

where $n = 0,1,2,3,4$. To be brief, the detailed deduction is omitted.

For example, if the core of the Galaxy rotates with equatorial speed $c_t = \boldsymbol{a}^4 c$ then the speed of gravitational interactions equal to $c_l = \boldsymbol{a}^{-4} c$. It coincides with the experimental value of the speed of gravitational interactions in the Galaxy determined according to data on co-rotational motion of the disc of the Galaxy on the distances till 10.3 kpc [8,9]. It is possible to describe co-rotational movement of the Galaxy as the solid state object only assuming that the retarding of interaction between the core and other stars of the Galaxy is absent. In this picture the Sun is retarded only on one itsr radius relatively to the vector of exchange interaction with the Galaxy core. So the Sun is sufficiently connected with the core of the Galaxy.

Besides that, the model proposed describes two maximums in the velocity spectrum of the stars. It was shown that these maximums are connected only with fields of electrons and protons, which are the most part of mass of the Galaxy. Theoretical estimations of Dubrovsky gives equal result for the speed of gravitational interaction (near $10^9 c$) [10] and it is closed to the Laplace estimation ($> 7 \cdot 10^6 c$) [2].

Interaction of cosmic objects is at different scales. So relative velocities of them will be also different. For example, the fastest planet of the Solar system is the Mercury. So it has the minimal speed of gravitational interaction. Let us express the orbit velocity of the Mercury through the radius-vector of its orbit by the formula

$$c_t = \boldsymbol{a}^2 c \, div \, \mathbf{R}_1 = 3\boldsymbol{a}^2 c. \qquad (10)$$

In the case the speed of its gravitational interactions is

$$c_l = \boldsymbol{a}^{-2} c = 1.878 \cdot 10^4 c. \qquad (12)$$

This value is a little less than Laplace had got. But this speed is critical in the Solar system when the motion is stable. From this point of view one may consider the Solar system as macrocondensate. Of course, velocities of gravitational interaction for other bodies of the Solar system will be higher because they move with more little speeds. For example, the speed of gravitational interaction for the Pluto is $1.89 \cdot 10^5 ñ.$

Thew other important question is about nature of inertial mass. There are no physical theories interpreting the physical meaning of the inertial mass. The model proposed give chance to explain it. The delay of gravitational field relatively the centre of mass plays the role of inertia. The elastic medium – gravitational field of the object – is weighed down under action of the external force accelerating the object. In this case the negative sign of counteracting force is appeared automatically. It is important that the gravitational field of the object will squeeze round it when its speed will increase. As the consequence, the decreasing of observed inertial mass should be found. It would be of very interest to test this experimentally.

In conclusion let us note about one more serious problem of gravity. There were a lot of experiments to define the Newton's gravity law more exactly last years. It was established trustworthy that the value of gravitational constant depends on distance between objects [11]. Thus, changing distance between objects from some centimeters till some meters $G$ increases on 1.1%. This effect is interpreted with the so-called 'fifth force'. In our mind, it is not necessary to attract the 'fifth force'. Let us assume that the dynamical part in the gravity law (6) is valid also at small distances taking the speed of heat motion of molecules of the bodies. Therefore, to calculate parameters of planet orbits one should take $G$ with taking into account large distances (Table 1. i.2.) [9].

**Conclusion**

Basic problems of current quantum relativiatic picture of the world can be solved by connecting gravity and electromagnetism. For this purpose basic ideas of Newton, Laplace, Mossotti, Maxwell, Heavisude were restored. Equations of de Broglie have got a new physical sense. Principles of relativity were also reconsidered. We had to refuse from geometrophysical approach in general relativity and to consider the set of Lorentz transformations in special relativity taking into account relative motion of the objects and their phase state. In the model proposed matter in the absolute vacuum defines the structure of the space. Newton had the continuous and linear model of space. The Einstein model of space is continuous and nonlinear. In the model proposed the space is discrete and nonlinear. Description of gravitational field in this scheme is based on superposition of two-dimensional objects: layers-discs. In such presentation interaction between objects is by waves moving along the layers of the space. To describe these waves linear second order differential equations are used. They gives that the speed of gravitational interaction is function of relative motion of the objects. At the limited speed of motion of condensed objects $\boldsymbol{a}^4 c$ the speed penetration of gravitational interaction along the axes of the space is $\boldsymbol{a}^{-4} c = 3.526 \cdot 10^8 c$.

Theoretical calculations are the most of cases are coinciding with experimental ones. Analysis of equations shows that the macroworld is quantized like macroworld.

**Acknowledgements**: I am very thankful to the personnel of the Consulting-Centre "Nanobiology" and Atomic and Molecular Engineering Laboratory for support and co-operation for many years.

Table 1

## Summary of formulas of the quantum astronomy

| N | Title | Theoretical formula | Theoretical value | Experimental value | Ref |
|---|---|---|---|---|---|
| | | *The autor* | | *Handbook* | |
| | | ***Fundamental constants*** | | | |
| 1 | Constant of fine structure | $a^{-1} = \sqrt{\left(\frac{N_a}{2p}\right)^2 + 1}$, $N_a = 861$ | 137.03605472… | 137.0360(2) | 9,8 |
| 2 | Gravitational constant on long distances | $G_\infty = \frac{e^2}{2p\varepsilon_0} \left(\frac{a^8}{4p(m_p + m_e)}\right)^2$ | 6.745991·$10^{-11}$ m³/rg·s² | (6.746±0.0024)·$10^{-11}$ m³/rg·s² | 11 |
| | | ***Solar system*** | | | |
| 3 | Everage orbit speed of the Mercury | $v_1 = 3a^2 c$ | 47.89307 km/s | 47.89 km/s | 8,9 |
| 4 | Maximum value of the large semiaxis of the Mercury orbit | $R_1 = \frac{h}{a^{12} m_p c} = \frac{z_0 e^2}{2a^{13} m_p c}$ | 5.795·$10^{10}$ m | 5.791·$10^{10}$ m | 8,9 |
| 5 | Maximum value of the large semiaxis of the Jupiter orbit | $R_5 = \frac{h}{a^{11} m_e c} = \frac{z_0 e^2}{2a^{12} m_e c}$ | 7.7647·$10^{11}$ m | 7.783·$10^{11}$ m | 8,9 |
| 6 | Ratio of maximum values of the large semiaxises of the Jupiter and Mercury | $\frac{R_5}{R_1} = a \frac{m_p}{m_e}$ | 13.3987 | 13.442 | 8,9 |
| | | ***Sun*** | | | |
| 7 | The first cosmic speed for the Sun | $v_{\odot I} = \frac{ac}{\sqrt{8p}}$ | 436.381 km/s | 436,78 km/s | 8,9 |
| 8 | Temperature in the centre of the Sun | $T_\odot = \frac{m_e v_{\odot I}^2}{2k} = \frac{m_e}{2k}\left(\frac{ac}{\sqrt{8p}}\right)^2$ | 6282.1K | 6270.0K | 8,9 |
| 9 | Period of longitudinal seismic waves on the surface of the Sun | $t_1 = 2p R_\odot \left(1 - a^{\frac{2}{3}}\right) v_{\odot I}^{-1}$ | 160.43 min | 160.01 min | 9, 12 |
| 10 | Period of transversal seismic waves on the surface of the Sun | $t_2 = 5a^{\frac{2}{3}} R_\odot v_{\odot I}^{-1}$ | 5.00 min | 5.00 min | 9, 12 |
| 11 | Equatorial speed of the Sun rotation | $v_\odot = a^2 c / 8$ | 1.995525 km/s | 1.9968 km/s | 8,9 |
| 12 | Period of rotation of the Sun round its own axis | $P_\odot = \frac{2p R_\odot}{v_\odot} = \frac{16p R_\odot}{a^2 c}$ | 25.364 days | 25.38 days | 8,9 |
| | | ***Planets*** | | | |
| 13 | Equatorial speed of the Earth rotation | $v_\oplus = 4a^3 c$ | 465.981 m/s | 465.10 m/s | 8,9 |
| 14 | Jupiter radius | $r_5 = \frac{GM_5 N_a}{4} \cdot \left(\frac{ac}{\sqrt{4p}}\right)^{-2}$ | 7.16326·$10^4$ km | 7.16326·$10^4$ km | 8,9 |
| 15 | Equatorial speed of rotation of the Jupiter surface | $v_5 = 2p \frac{a^2 c}{8}$ | 12.5383 km/s | 12.55 km/s | 8,9 |
| *Galaxy* | | | | | |
| 16 | Maximum speed of stars round the Galaxy core | $V_1 = \dfrac{ac}{8}$ | 273.46 km/s | 273 km/s | 8,9 |
| 17 | Maximim relative speed of closed stars | $V_2 = a^2 c$ | 15.964 km/s | 15.5 km/s | 8,9 |
| 18 | Distance to the first maximum in the spectrum of star speeds relatively to the Galaxy core | $R_{Gp} = \dfrac{R_1}{a^4} = \dfrac{h}{a^{16} m_p c}$ | $2.043 \cdot 10^{19}$ m = =0.6622 kpc | 0.5-0.8 kpc | 8,9 |
| 19 | Distance to the second maximum in the spectrum of star speeds relatively to the Galaxy core | $R_{Ge} = \dfrac{R_5}{a^4} = \dfrac{h}{a^{15} m_e c}$ | $2.738 \cdot 10^{20}$ m =8.87 kpc | 8 – 10 kpc | 15 |
| *Methagalaxy* | | | | | |
| 20 | Habble constant | $H_0 = \dfrac{a^{18} m_e c^2}{\hbar}$ | 82.489 kms$^{-1}$Mpc$^{-1}$ | 50÷100 kms$^{-1}$Mpc$^{-1}$ | 8,9 |
| 21 | Radius of Methagalaxy | $R_M = \dfrac{c}{H_0} = \dfrac{\hbar}{a^{18} m_e c}$ | $1.1214 \cdot 10^{23}$ km | ? | 8,9 |

$\alpha$ is the constant of fine structure or longitudinal quantum number; $e$ is the elementary charge; $\hbar = \dfrac{h}{2\pi}$ Planck constant; $c$ is the speed of light; $m_e$ is the mass of an electron; $m_p$ is the mass of a proton; $k$ is the Boltsman constant, $G$ is gravitational constant, $z_0$- the vacuum wave resistance.